\documentclass{PoS}
\newcommand{\Slash}[1]{{\ooalign{\hfil/\hfil\crcr$#1$}}}
\title{In-medium $\eta'$ mass and $\eta'N$ interaction in vacuum in the
linear sigma model\footnote{Report No. : KUNS-2473}}

\ShortTitle{In-medium $\eta'$ mass and $\eta'N$ two-body interaction}

\author{\speaker{Shuntaro Sakai}\\
        Department of Physics, Kyoto University, Kitashirakawa-Oiwakecho,
        Kyoto 606-8502, Japan\\
        E-mail: \email{s.sakai@ruby.scphys.kyoto-u.ac.jp}}

\author{Daisuke Jido\\
        Department of Physics, Tokyo Metropolitan University, Hachioji,
        Tokyo 192-0397, Japan\\
        E-mail: \email{jido@tmu.ac.jp}}

\abstract{We investigate the $\eta'N$ two-body interaction
  in the context of the $\eta'$ meson mass modification in the nuclear medium.
  It has been argued in several articles that the masses of $\eta'$ and
  the other pseudoscalar mesons ($\pi$, $K$, $\eta$) should degenerate in the
  chiral-symmetric phase.
  It is expected that the reduction of the mass difference between
  $\eta$ and $\eta'$ would take place in the
  nuclear matter if one assumes that the decrease of the quark
  condensate at the normal nuclear density occurs with
  partial restoration of chiral symmetry.
At low density, the in-medium self-energy giving the mass modification
  by the medium effect can be obtained by the $\eta'N$ two-body $T$ matrix.
  Thus, we also estimate the $\eta'N$ interaction strength
  in vacuum with the linear sigma model which involves the effect of
  partial restoration of chiral symmetry.
  In the view of the linear sigma model, we find that
  the $\eta'N$ interaction is attractive and generated through the sigma
  meson exchange.
  We expect that the interaction is enough strong and for the existence
  of a bound state of the $\eta'N$ system.}

\FullConference{XV International Conference on Hadron Spectroscopy-Hadron 2013\\
		4-8 November 2013\\
		Nara, Japan }

\usepackage{amsmath}	
\begin{document}
\section{Introduction}
The mass spectrum of hadrons reflects the symmetry of its fundamental
theory, Quantum Chromodynamics (QCD).
If one considers simply, nine Nambu--Goldstone (NG) bosons would
appear according to the NG theorem if one assumes the axial part of
U(3)$_L\times$U(3)$_R$ is spontaneously broken down to U(3)$_V$.
Nevertheless, when one sees the pseudoscalar meson
masses, one cannot find the singlet pseudoscalar meson around the
pion mass \cite{Weinberg1975}.
Taking account of the U$_A$(1) anomaly, we can regard the U$_A$(1)
symmetry as broken explicitly and the $\eta'$ mass can be expained with
the U$_A$(1) anomaly \cite{Witten1979}.

Other than the U$_A$(1) anomaly, chiral symmetry breaking
also plays an important role for the generation of the $\eta'$ mass.
As discussed in Refs. \cite{Cohen1996}, the
pseudoscalar singlet meson and octet meson degenerate when chiral
symmetry is restored, e.g. at high temperature or high density, in the
chiral limit even if the U$_A$(1) symmetry is broken explicitly.
Taking account of the degeneracy of the pseudoscalar flavor-singlet and
octet meson in the chiral restored phase and the partial restoration of
chiral symmetry, we expect the reduction of the $\eta'$ mass in the
nuclear matter.
Concerning the chiral restoration in the nuclear matter, the 35\%
reduction of the quark condensate is suggested from the analysis of the
experimental data \cite{Suzuki2004}.
Recent experiments have suggested that the $\eta '$-nucleus optical
potential could be attractive with certain strength and could hava a smaller
imaginary part \cite{Nanova2013}.
We can interpret the $\eta'$ mass reduction in the nuclear matter as the
attractive potential of $\eta'$ in the nuclear matter.
The existence of $\eta'$-mesic nuclei is suggested theoretically
\cite{Saito2007} and the experimental attempt to observe
the $\eta'$-mesic nuclei is discussed \cite{Itahashi2012}.
However, it is not known well whether the interaction between $\eta'$
and nucleon is attractive or repulsive despite the existence of some
experimental data \cite{Moskal1}.
Such a poor knowledge of the $\eta'N$ interaction makes it difficult to
analyze the $\eta'$ properties in the nuclear matter.

In the below, we study the $\eta'N$ two-body interaction with the linear
sigma model as a chiral effective model.
In the construction of the model, we assume the 35\% reduction of the
quark condensate.
In addition to the $\eta'N$ interaction, we calculate the in-medium
$\eta'$ mass which is expected to reduce in the nuclear matter.
The detail of this work is shown in Ref. \cite{Sakai2013}
\section{Method}
For the calculation, we use the SU(3) linear sigma model as a chiral effective
model \cite{Schechter1971,Renaghan2000}.
The linear sigma model can describe both the chiral restored phase and
the spontaneously broken phase.
To describe the $\eta'N$ interaction, we introduce the
nucleon degree of freedom explicitly based on the chiral symmetry.
The Lagrangian is given as
\begin{eqnarray}
 \mathcal{L}&=&\frac{1}{2}{\rm tr}\partial_\mu M\partial^\mu
  M^\dagger-\frac{\mu^2}{2}{\rm tr}MM^\dagger-\frac{\lambda}{4}{\rm
  tr}(MM^\dagger)^2-\frac{\lambda'}{4}\left({\rm tr}MM^\dagger\right)^2\nonumber\\
 & &+A{\rm tr}\left(\chi
	       M^\dagger+M\chi^\dagger\right)+\sqrt{3}B\left(\det M
	       +\det M^\dagger \right) \nonumber\\
 & &+\bar{N}i\Slash{\partial}N-g\bar{N}\left(\frac{\sigma_0}{\sqrt{3}}+\frac
{\sigma_8}{\sqrt{6}}+i\gamma_5\frac{\vec{\tau}\cdot\vec{\pi}}{\sqrt{2}}+i\gamma_5\frac{\eta_0}{\sqrt{3}}+i\gamma_5\frac{\eta_8}{\sqrt{6}}\right)N.
\end{eqnarray}
Here, the meson field, the nucleon field, and the quark mass are given,
respectively, by
\begin{eqnarray}
 M=\sum_{a=0}^8\frac{\sigma_a\lambda_a}{\sqrt{2}}+i\sum_{a=0}^8\frac{\pi_a\lambda_a}{\sqrt{2}},\ N=
 (p,n)^{\rm t} ,\ \chi={\rm diag}(m_q, m_q, m_s).
\end{eqnarray}
The Lagrangian is constructed to possess the same global symmetry as QCD.
The term proportional to $A$ expresses the effect of the current quark
mass.
Here, $\chi$ corresponds to the current quark mass and we assume the
isospin symmetry, $m_q=m_u=m_d$, and introduce the SU(3) flavor symmetry
breaking with $m_q\neq m_s$.
The term proportional to $B$ represents the effect of the U$_A$(1) anomaly
and this term is not invariant under the U$_A$(1) transformation.

The Lagrangian contains 6 free parameters which cannot be fixed
from the symmetry.
We fix these parameters using the observed meson masses, the meson
decay constants and the fact that the 35\% reduction of the quark
condensate at the normal nuclear density.
For the calculation of the in-medium quantities, i.e. the meson
masses, we introduce the effect of the symmetric nuclear matter with the
mean field approximation of nucleon.

In the linear sigma model, the vacuum expectation value of the sigma
field $\left<\sigma_0\right>$ is an order parameter of the chiral symmetry
breaking.
Now, we have non-zero $\left<\sigma_8\right>$ due to the explicit flavor
symmetry breaking.
We determine $\left<\sigma_0\right>$ and $\left<\sigma_8\right>$ to
minimize the effective potential.
\section{Results}
\subsection{In-medium meson mass}
\begin{figure}
\begin{minipage}{0.45\hsize}
\begin{center}
  \includegraphics[width=5cm]{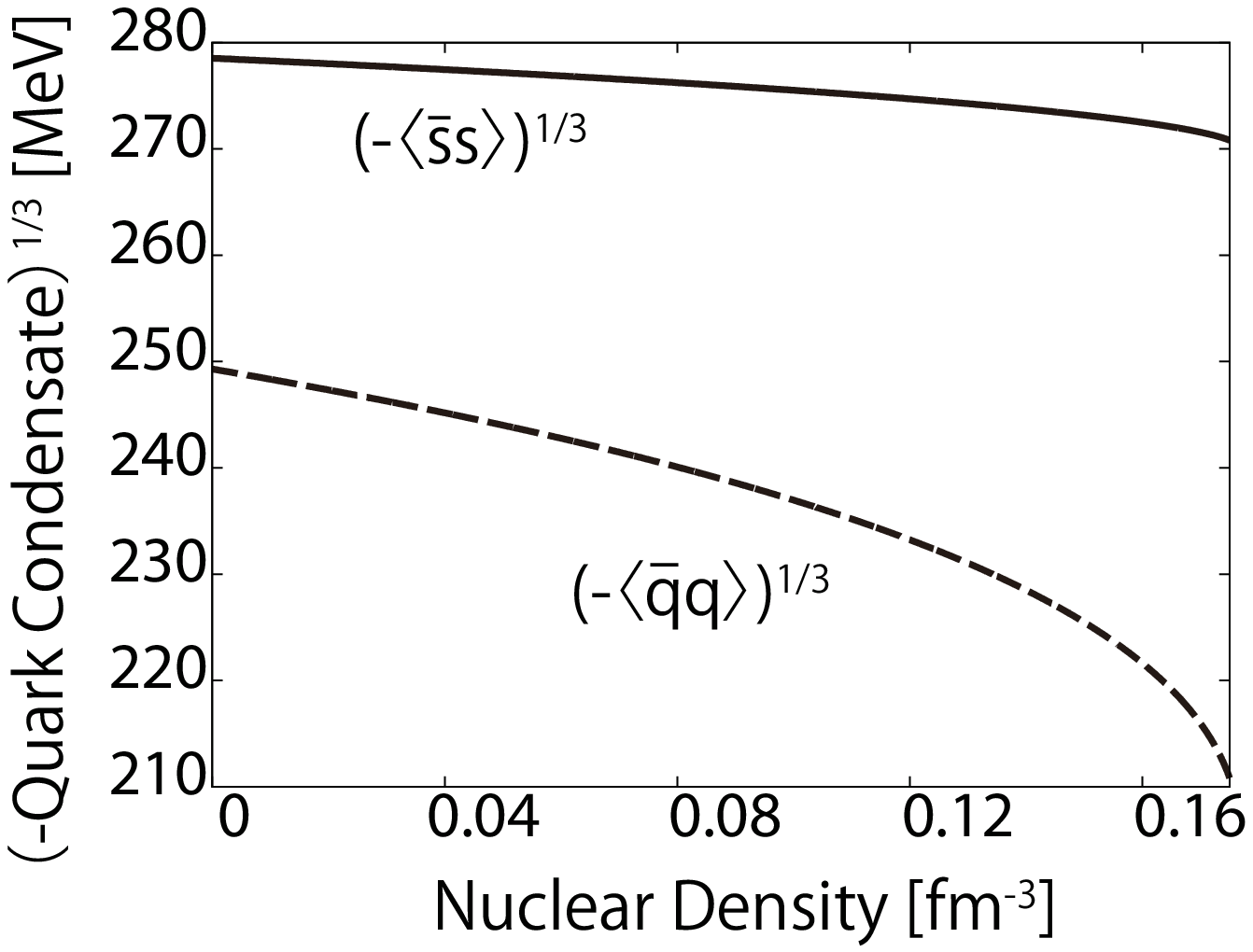}
\end{center}
\caption{In-medium quark condensate. The dashed and solid lines
 represent the in-medium $u$, $d$ and $s$ quark condensates,
 respectively.}
\label{quarkcondensate}
\end{minipage}
\begin{minipage}{0.1\hsize} 
\end{minipage}
\begin{minipage}{0.45\hsize}
\begin{center}
  \includegraphics[width=4.5cm]{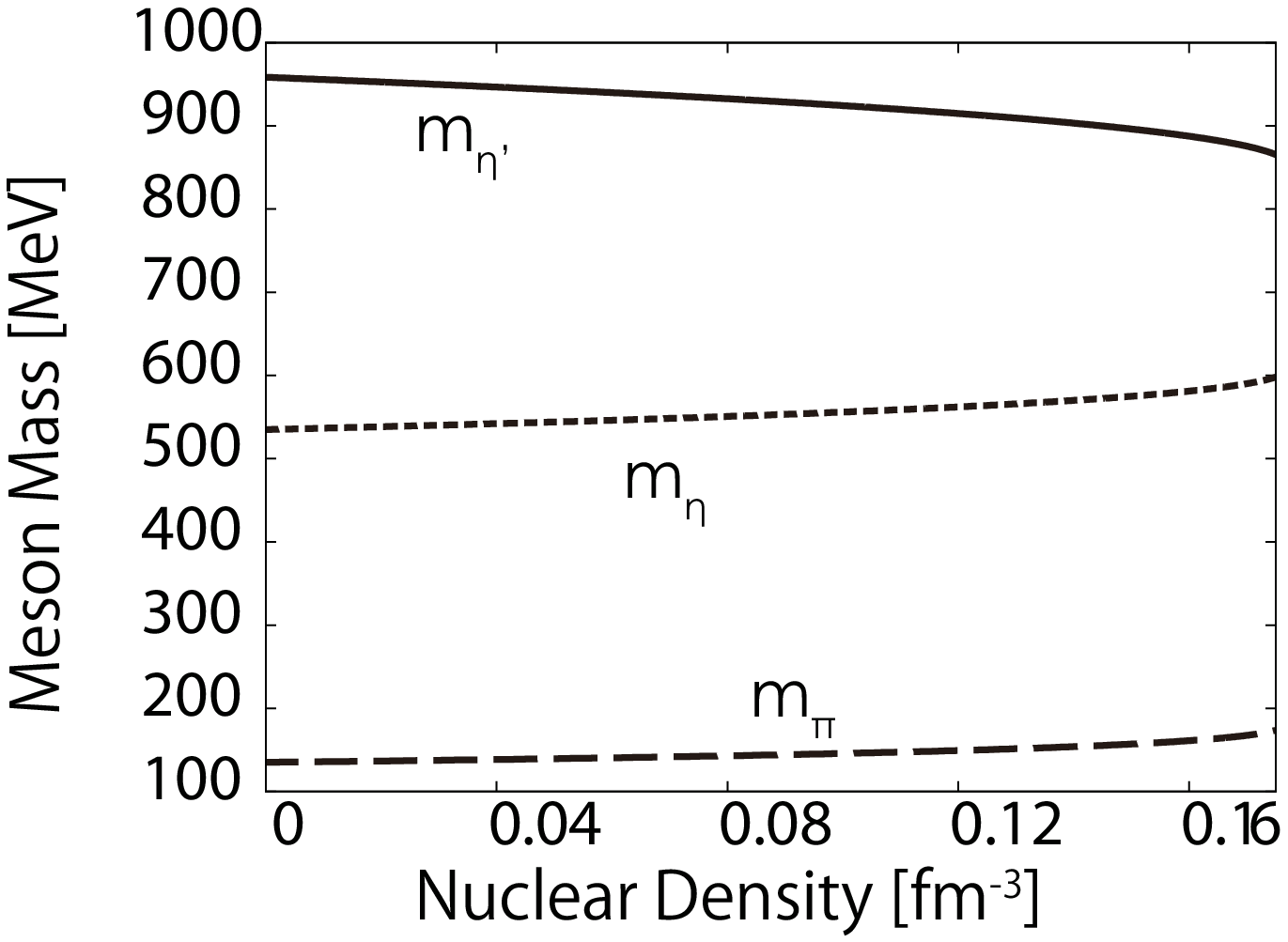}
\end{center}
\caption{In-medium meson mass. The solid, dotted and dashed lines
 represent the $\eta'$, $\eta$ and $\pi$ masses, respectively.}
\label{mesonmass}
\end{minipage}
 \end{figure}
First, we show the in-medium quark condensate in Fig. \ref{quarkcondensate}.
As mentioned above, we assume the 35\% reduction of the quark
condensate, so the value of the $u, d$ quark condensate at the normal
nuclear density is the input value.
Here, we have assumed the isospin symmetry, so the $u$, $d$ quark
condensates coinside.
Next, we discuss the in-medium meson mass.
The in-medium self-energy of the mesons comes from the diagrams shown in
Fig. \ref{loop}.
Diagram ($a$) in Fig.\ref{loop} comes from the
nucleon mean field, while ($b$) and ($c$) come from the particle-hole
excitation and the crossed channel of the particle-hole excitation,
respectively.
In the chiral limit, the $\eta'$ mass can be written as
\begin{eqnarray}
 m_{\eta'}^2=6B\left<\sigma_0\right>.\label{etaprimemass}
\end{eqnarray}
$B$ is the coefficient of the determinant term and represents the effect of
the U$_A$(1) anomaly.
From this expression, one can find the necessity of the U$_A$(1) anomaly and
the chiral symmetry breaking for the generation of the $\eta'$ mass, and
the restoration of chiral symmetry, the reduction of
$\left<\sigma_0\right>$, leads to the reduction of the $\eta'$ mass.
The calculated in-medium meson masses are shown in Fig. \ref{mesonmass}
From the calculation, the $\eta'$ mass reduces about 80MeV and the
$\eta$ mass enhances about 50MeV at the normal nuclear density.
The mass difference between the $\eta$ and $\eta'$ mass reduces about
130MeV, going toward the degeneracy of $\eta$ and $\eta'$, as we have expected.
\subsection{$\eta' N$ 2-body interaction}
\begin{figure}
\begin{minipage}{0.45\hsize}
\centering
\includegraphics[width=6cm]{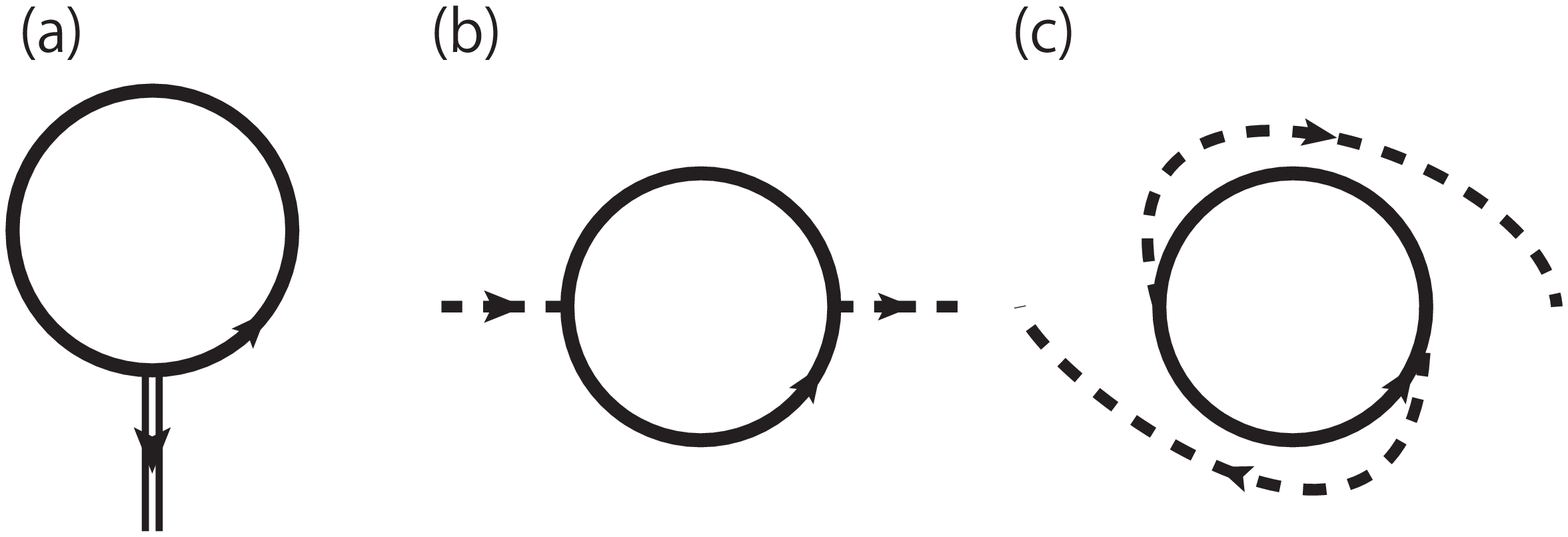}
\caption{The diagrams contributing to the in-medium meson
 self-energy.}
\label{loop}
 \end{minipage}
\begin{minipage}{0.1\hsize} 
\end{minipage}
\begin{minipage}{0.45\hsize}
 \begin{center}
  \includegraphics[width=6cm]{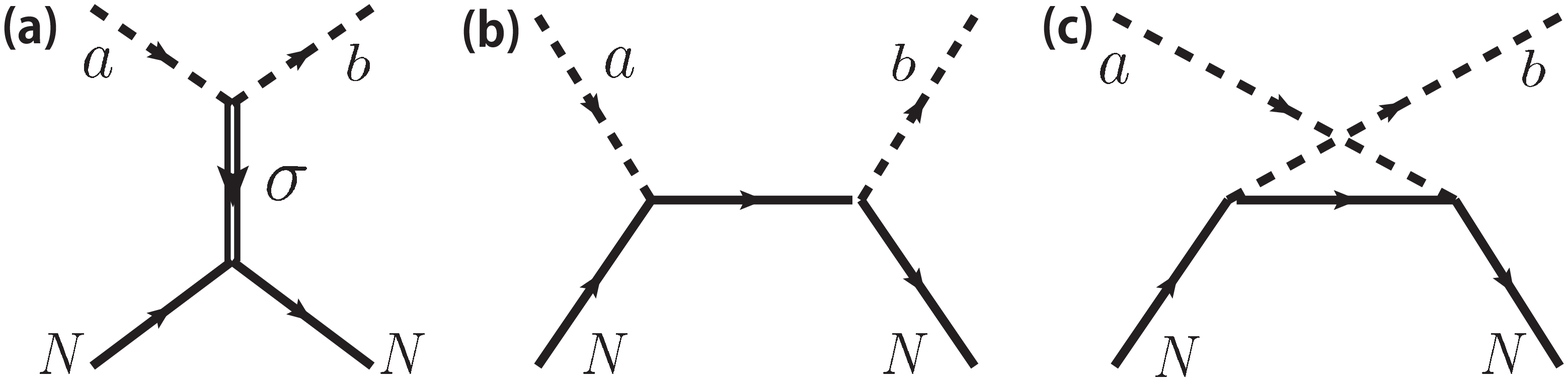}
 \end{center}
\caption{The diagrams which contribute to the $\eta'N$ interaction}
\label{diag}
\end{minipage}
\end{figure}
Here, we show the $\eta'N$ 2-body interaction evaluated with the same linear
sigma model.
In the linear sigma model, the diagrams shown in Fig. \ref{diag} contribute
to the $\eta'N$ interaction.
The diagram ($a$) in the Fig. \ref{diag} shows the contribution from the sigma
meson exchange, while diagram ($b$) and ($c$) in Fig. \ref{diag} are
the contributions from the Born term which contain the nucleon
intermediate state.
From these diagrams, we have obtained the low-energy $\eta'N$ 2-body
interaction $V_{\eta'N}$ in the chiral limit as
\begin{equation}
 V_{\eta'N}=-\frac{6gB}{\sqrt{3}m_{\sigma_0}^2}\label{etaprimen}.
\end{equation}
Substituting the value of the fixed parameters, we find that the
$\eta'N$ interaction is comparably strong to the $\bar{K}N$ system.
In the $\bar{K}N$ system, a bound state, $\Lambda(1405)$, exists due to
the strong $\bar{K}N$ attraction.
With the analogy to $\Lambda(1405)$, we expect the existence of the
$\eta'N$ bound state.
For the investigation of the possibility of the bound state, we analyzed
the T-matrix of the $\eta'N$ system because the bound state appears as
the pole of the T-matrix.
We obtain the T-matrix with solving the single-channel Lippman-Schwinger
equation,
\begin{eqnarray}
 T&=&V+VGT.
\end{eqnarray}
Here, $G$ is the loop integral of $\eta'$ and nucleon,
\begin{eqnarray}
 G(W)&=&2m_N\int\frac{d^4q}{(2\pi)^4}\frac{1}{(P-q)^2-m_N^2+i\epsilon}\frac{1}{q^2-m_N^2+i\epsilon},
\end{eqnarray}
$P=(W,{\bf 0})$ is the 4-momentum of the $\eta'N$ in the center of mass system.
As the interaction kernel $V$, we use the $\eta'N$ interaction obtained
with linear sigma model shown in Eq. (\ref{etaprimen}).
Now, the interaction kernel is momentum-independent, so the equation can
be solved with the algebraic way,
\begin{eqnarray}
 T(W)=\frac{1}{V^{-1}-G(W)}.
\end{eqnarray}
The obtained T matrix contains a divergence in the loop integral $G$.
Here, we regulate the divergence with dimensional regularization and we fix
the subtraction constant with the natural renormalization scheme, which
exclude the other dynamics than $\eta'$ and $N$ \cite{Hyodo2008}.
We have found a $\eta'N$ bound state as a pole of the obtained T matrix.
The binding energy is 6.2 MeV, the scattering length is -2.7 fm and
the effective range is 0.25 fm.
The scattering length is the repulsive sign in our notation.
Here, we note that the obtained $\eta'N$ scattering length is somewhat
larger value compared to the value suggested in Ref. \cite{Moskal1}.
\section{Conclusion}
In this paper, we have calculated the in-medium meson mass and the
$\eta'N$ 2-body interaction with the SU(3) linear sigma model.
The medium effect is introduced as one nucleon loop for the
calculation of the in-medium meson mass.
We have obtained about 80 MeV reduction of the $\eta'$ mass and 130 MeV
decrease of the mass difference between $\eta$ and $\eta'$. 
Concerning the $\eta'N$ two-body interaction, we have found the strong
attraction of $\eta'N$ comparable to the $\bar{K}N$ system.
The $\eta'N$ interaction obtained from the linear sigma model is
provided from the sigma meson exchange.
This is a different character from that of the ordinary NG boson, the
Weinberg--Tomozawa interaction which is energy dependent.
With the analogy of $\Lambda(1405)$, we have investigated the
possibility of the $\eta'N$ bound state.
As a result, we found a $\eta'N$ bound state with the binding energy
6.2 MeV and the scattering length -2.7 fm.
The coupling between $\sigma_0$ and $\eta'$ is necessary for the
generation of the $\eta'$ mass and the $\eta'$-$\sigma_0$ coupling leads
to the attraction through the sigma meson exchange.
\acknowledgments
S.S. is a JSPS Fellow and appreciates the support by a
JSPS Grant-in-Aid (No. 25-1879). This work was partially
supported by Grants-in-Aid for Scientific Research from
MEXT and JSPS (No. 25400254 and No. 24540274).


\begin{thebibliography}{99}
\bibitem{Weinberg1975}
S.~Weinberg, Phys. Rev. {\bf D11} (1975) 3594.
\bibitem{Witten1979}
E.~Witten, Nucl. Phys. {\bf B156} (1979) 269; G.~Veneziano, Nucl. Phys. {\bf B159} (1979) 213.
\bibitem{Cohen1996}
T.D.~Cohen, Phys. Rev. {\bf D54} (1996) 1867; S.H.~Lee, T.~Hatsuda,
	Phys. Rev {\bf D54} (1996) 1871; D.~Jido, {\it et al}., Phys. Rev. {\bf C85} (2012) 032201.
\bibitem{Suzuki2004}
K.~Suzuki, {\it et al}., Phys. Rev. Lett. {\bf 92} (2004) 72302; E.~Friedman, {\it et al}., Phys. Rev. Lett. {\bf 93} (2004) 122302.
\bibitem{Nanova2013}
M.~Nanova, {\it et al.}, Phys. Lett. {\bf B710} (2012) 600; {\it ibid}
	{\bf B727} (2013) 417.
\bibitem{Saito2007}
H.~Nagahiro, S.~Hirenzaki, Phys. Rev, Lett. {\bf 94} (2005) 232503; K.~Saito, {\it et al}., Prog. Part. Nucl. Phys. {\bf 58} (2007) 1.
\bibitem{Itahashi2012}
K.~Itahashi, {\it et al}., Prog. Theor. Phys. {\bf 128} (2012) 601.
\bibitem{Moskal1}
P.~Moskal, {\it et al}., Phys. Lett. {\bf B474} (2000) 416; P.~Moskal, {\it et al}., Phys. Lett. {\bf B482} (2000) 356.
\bibitem{Sakai2013}
S.~Sakai, D.~Jido, Phys. Rev. {\bf C88} (2013) 064906.
\bibitem{Schechter1971}
J.~Schechter, Y.~Ueda, Phys. Rev. {\bf D3} (1971) 168.
\bibitem{Renaghan2000}
J.T.~Renaghan, {\it et al}., Phys. Rev. {\bf D62} (2000) 085008.
\bibitem{Hyodo2008}
T.~Hyodo, D.~Jido, A.~Hosaka, Phys. Rev. {\bf C78} (2008) 025203.

\end{thebibliography}
\end{document}